\documentclass[manuscript,screen]{acmart}
\AtBeginDocument{%
  \providecommand\BibTeX{{%
    \normalfont B\kern-0.5em{\scshape i\kern-0.25em b}\kern-0.8em\TeX}}}

\setcopyright{acmlicensed}
\copyrightyear{2024}
\acmYear{2024}
\acmDOI{XXXXXXX}

\begin{document}

\title{DinAR: Augmenting Reality for Sustainable Dining}

\author{MJ Johns}
\email{mljohns@ucsc.edu}
\orcid{0009-0002-4016-2517}
\affiliation{%
  \institution{University of California Santa Cruz}
  \streetaddress{1156 High St}
  \city{Santa Cruz}
  \state{California}
  \country{USA}
  \postcode{95064}
}
\author{Eunsol Sol Choi}
\email{echoi33@ucsc.edu}
\orcid{0000-0002-4397-5751}
\affiliation{%
  \institution{University of California Santa Cruz}
  \streetaddress{1156 High St}
  \city{Santa Cruz}
  \state{California}
  \country{USA}
  \postcode{95064}
}
\author{Derusha Baskaran}
\email{dbaskara@ucsc.edu}
\orcid{0000-0002-2056-3025}
\affiliation{%
  \institution{University of California Santa Cruz}
  \streetaddress{1156 High St}
  \city{Santa Cruz}
  \state{California}
  \country{USA}
  \postcode{95064}
}

\begin{abstract}
Sustainable food is among the many challenges associated with climate change. The resources required to grow or gather the food and the distance it travels to reach the consumer are two key factors of an ingredient’s sustainability. Food that is grown locally and is currently “in-season” will have a lower carbon footprint, but when dining out these details unfortunately may not affect one's ordering preferences. We introduce DinAR as an immersive experience to make this information more accessible and to encourage better dining choices through friendly competition with a leaderboard of sustainability scores. Our study measures the effectiveness of immersive AR experiences on impacting consumer preferences towards sustainability. 

\end{abstract}



\begin{teaserfigure}
  \includegraphics[width=\textwidth]{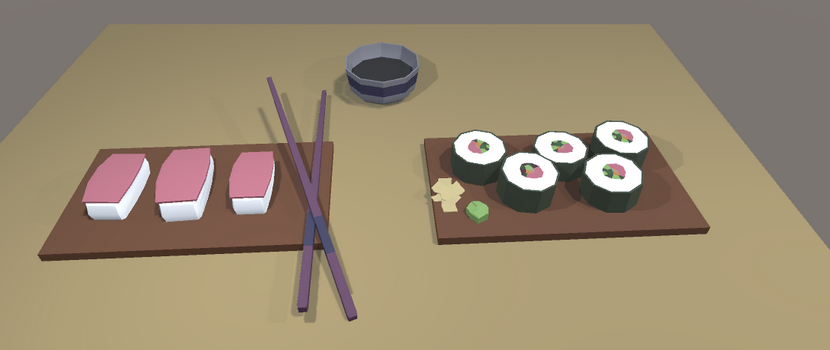}
  \caption{Making sustainable choices easier when dining out}
  \Description{A 3D model of a sushi dinner}
  \label{fig:teaser}
\end{teaserfigure}

\maketitle

\section{Introduction}
Climate change is a growing problem that our society faces today and seemingly so for the decades to come due to human activities and behaviors \cite{myers_consensus_2021, lynas_greater_2021}. Food sustainability, however, is one solution to mitigate a part of climate change and enhance the overall well-being of our society \cite{smith_climate_2013,gomez-zavaglia_corrigendum_2020}. Food sustainability includes, but is not limited to, concepts such as local sourcing and seasonal produce \cite{morawicki_food_2018}. Local sourcing is when people purchase produce from local farms, markets, and businesses that are in or near their community \cite{enthoven_local_2021}. Seasonal produce refers to food that is currently in its peak time period for harvesting – purchasing food within this time frame can reduce greenhouse gas emissions \cite{vargas_role_2021}. 

When purchasing groceries, one can take more time to consider or look up what produce is in season and how it came to their local area. Unfortunately, when dining out, many people do not make the same considerations and are unaware of their meals’ food sustainability, either due to limitations of time or availability of the information. In today’s digital era, many people have a mobile phone and often use it when eating a meal. Although people are already actively using their phones during dine-out meals, they still lack knowledge and awareness about food sustainability even when plenty of information is virtually within their reach. 

And so, nudging human behaviors towards more conscious choices in ordering food can help improve food sustainability while reducing our impact on climate change. Hence, we ask this as our research question: Can an interactive AR experience impact consumer food preferences toward more sustainable choices when dining out?

\section{Background}
\subsection{Food sustainability and human behavior}
Human behavior and choices can impact food sustainability; however, the information can be difficult to retrieve and parse. Some of the factors that impact a food’s sustainability are food sourcing (i.e., how far it needed to travel and by what means it was transported) and seasonal food (based on climates and seasons when produce grows well) \cite{morawicki_food_2018, enthoven_local_2021, vargas_role_2021}. 

With our app, we seek to support this greener decision-making when ordering a meal while dining out. In addition to using sustainability factors when making a decision about what to buy and eat, poor purchase planning also plays a factor by resulting in over-buying and food waste \cite{morawicki_food_2018}. We can also see this at a smaller scale when dining out when it is difficult to tell how much food will be included in an order. Our app also addresses this issue by offering a to-scale model of the meal to help understand the food portioning and allow for better decision-making. 

\subsection{Food Sustainability and AR}
AR and other immersive experiences offer promising opportunities for sustainability by presenting information in more accessible, relatable, and tangible ways \cite{han_designing_2022}. Nonetheless, there are some challenges, such as social acceptance and technological specifications, that arise with the implementation of AR in food sustainability \cite{jagtap_food_2021}.

Given that AR is a relatively new technology that does require a large battery and computing power from compatible devices, naturally, it may take some time for people to adopt and become more familiar with an AR experience while dining \cite{jagtap_food_2021,alzahrani_augmented_2020}. 

To overcome these challenges, we have designed our app to be easily accessible and socially engaging, as users simply scan a QR code from a menu to start this AR experience and can partake in friendly competition with others through a leaderboard (see Section 4: AR Food Interaction for more details on how our app works). Furthermore, the minimal yet informative design throughout our AR app ensures that this experience is not too technologically demanding on users’ AR-compatible mobile devices — especially since we encourage people to use our app at the beginning of their dining experience when making a meal choice. While our work is in the early stages, we seek to provide additional evidence in support of this effort through the development of our AR app, DinAR.

\section{System Overview}

\begin{figure}
    \centering
    \includegraphics[width=1\linewidth]{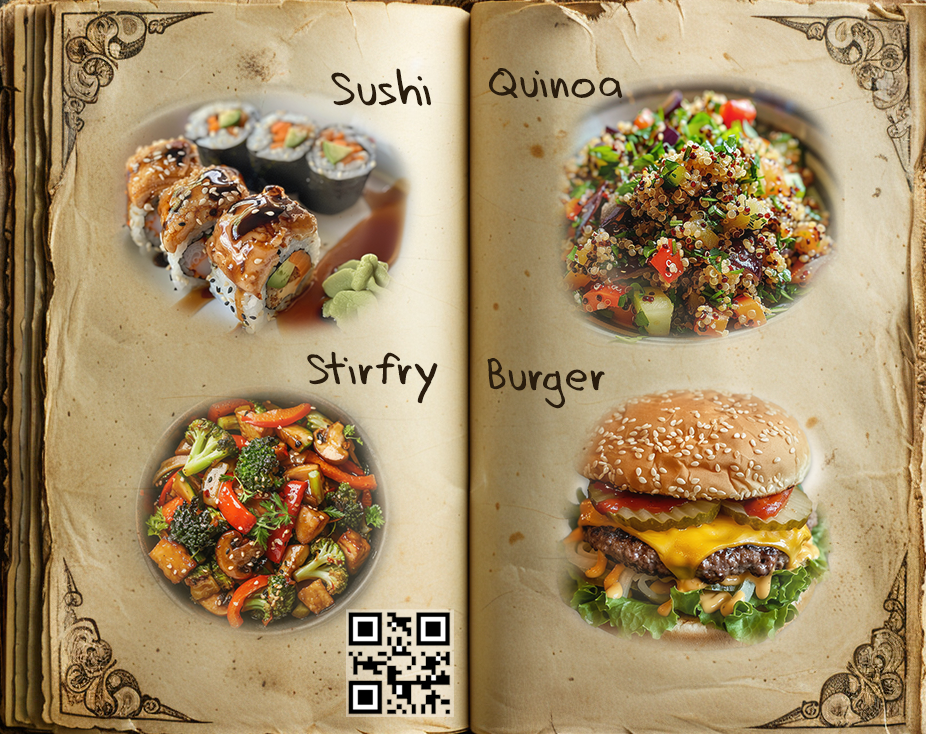}
    \caption{Playable AR demo, scan QR code to launch app, then aim at one of the dishes to see its sustainability score (images generated with MidJourney)}
    \label{fig:menu}
\end{figure}

We are building an interactive Augmented Reality (AR) mobile app that will enable diners at a restaurant to quickly and easily pull up sustainability information from the menu before deciding what to order. The app will overlay key details about each dish, including where each ingredient is sourced and whether it is in season, and (stretch goal) a 3D model showing the size of the dish to help reduce food waste from over-ordering. The app is playable by interacting with Figure \ref{fig:menu} using your phone. 

\section{AR-Food Interaction}

\begin{figure}[ht]
    \centering
    \includegraphics[width=0.5\linewidth]{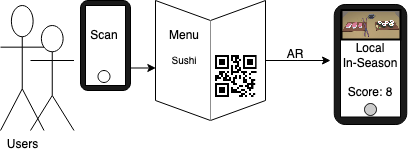}
    \caption{User flow example of how users will interact with our AR experience while dining out}
    \label{fig:diagram}
\end{figure}

The user (person eating at a restaurant) will use their phone to scan the QR code on the menu to launch the AR app. Then, while looking through their phone at the menu, they will see details overlaid on each menu item to help them make a sustainable choice. For these details shown, we plan to use the Environmental Working Group \footnote{https://www.ewg.org/foodscores/} (EWG)’s sustainable food scoring system, which combines three factors: nutrition, ingredient concerns, and degree of processing, to provide users with a sustainable score for each meal. This user flow can be seen in Figure \ref{fig:diagram}.

Beyond interacting individually with the app, we envision a leaderboard-style interaction \textit{with other humans} to compare sustainability scores based on their choices over time (such as a weekly leaderboard).

\begin{figure}[ht]
    \centering
    \includegraphics[width=.75\linewidth]{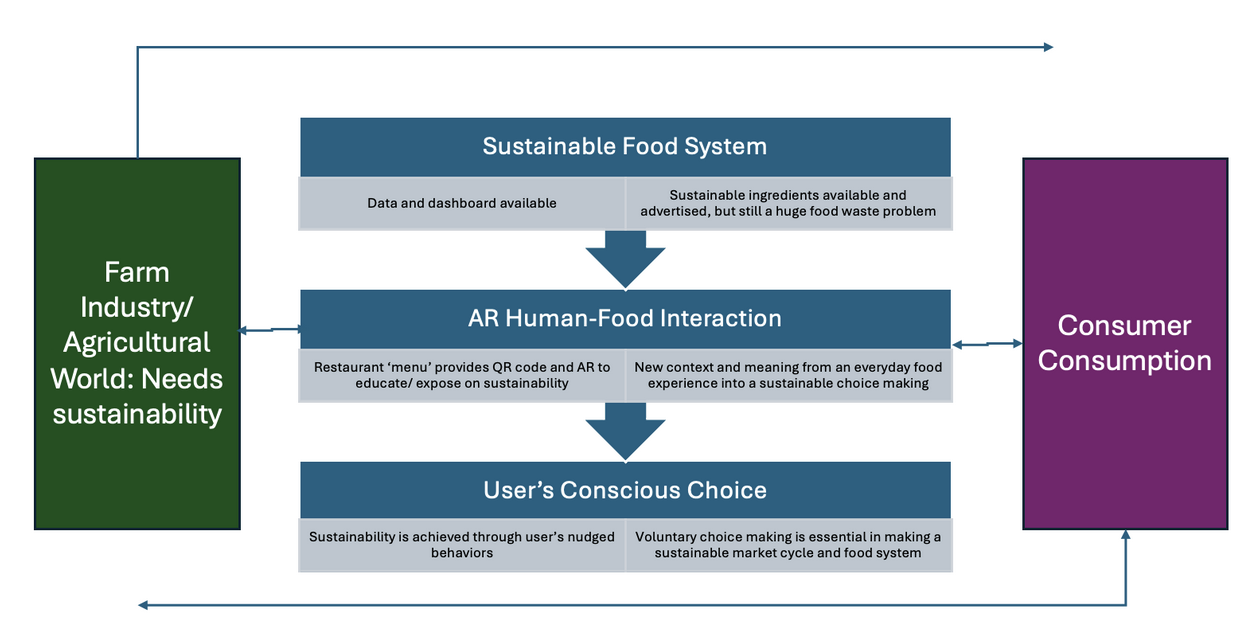}
    \caption{A proposed holistic framework for considering consumer behavior within the food system}
    \label{fig:framework}
\end{figure}

As part of our development of the AR experience, we propose a human-food interaction framework to consider a holistic approach to a sustainable food system that considers consumer food choices, as seen in Figure \ref{fig:framework}.

Currently, there exists a gap between the urgent need for an integrated food security system and unassisted choice-making processes among stakeholders in food systems \cite{horton_integrated_2016}. AR technology in restaurants can enable both an entertaining and intriguing experience \cite{han_designing_2022} and a new meaning-making from menu reading by sharing and mapping sustainable agri-food data. A nudging effect can be achieved from consumers consciously choosing their foods; the effect should be a virtuous cycle of sustainable food consumption and the benefited agri-food society. Collectively, the proposed framework is shown in the above diagram.

\section{Envisioned Evaluation}

Although this project is early in development, we plan to use the following approaches to evaluating the effectiveness of our system in changing opinions and behavior around sustainable dining choices:
\begin{itemize}

\item Pre- and Post-survey before and after using the prototype
\item A/B testing against “traditional” method of persuading towards sustainability (printed flier with details)
\item Long-term survey after a week of leaderboard competition with friends
\end{itemize}
We anticipate that this evaluation will provide evidence supporting using interactive AR experiences to encourage sustainability-focused behavior change.

\section{Conclusion}

Sustainable dining choices are easier to make when you have all of the information in an accessible (and entertaining) format. Our augmented reality approach with DinAR makes cumbersome information fun and engaging as an AR overlay on a traditional menu. With this interactive experience, we hope to encourage people to make sustainable choices when dining out.

\bibliographystyle{ACM-Reference-Format}
\bibliography{references}

\end{document}